\documentstyle[prl,aps]{revtex}
\begin{document}
\draft
\voffset=.5in
%@twocolumnfalse\endcsname

\title
{Cutoff dependence of the Casimir effect}

\author {C. R. Hagen\cite{Hagen}}

\address
{Department of Physics and Astronomy\\
University of Rochester\\
Rochester, N.Y. 14627}

\maketitle

\begin{abstract}
The problem of calculating the Casimir force on two conducting planes by means
of the stress tensor is examined.  The evaluation of this quantity is carried
out using an explicit regularization procedure which has its origin in the
underlying (2+1) dimensional Poincar\'{e} invariance of the system.  The force
between the planes is found to depend on the ratio of two independent
cutoff parameters, thereby rendering any prediction for the Casimir effect an
explicit function of the particular calculational scheme employed.  Similar
results are shown to obtain in the case of the conducting sphere.

\end{abstract}

\pacs {11.10Kk, 12.20.Ds}

%*
%\vskip2pc]

     In 1948 Casimir [1] first predicted that two infinite parallel plates in
vacuum would attract each other.  This remarkable result has its origin in the
zero point energy of the electromagnetic field.  While the latter is highly
divergent, the change associated with this quantity for specific plate
configurations has been found in numerous calculations to be finite as well as
cutoff dependent and thus in principle observable.  Early work to detect this
small effect [2] was characterized by relatively large experimental
uncertainties which left the issue in some doubt.  More recent efforts [3] have
provided quite remarkable data, but are based on a different geometry from that
of Casimir.  Since a rigorous theoretical calculation has never been carried
out for the latter configuration, there remains room for skepticism as to
whether the Casimir effect is as well established as is frequently asserted.

The most elementary calculation of the Casimir effect between two parallel
conducting planes located at $z=0$ and $z=a$ employs a mode summation in the
framework of a regularization which depends only on the frequency
${\omega}_k = [{\bf k}^2 + ({n\pi\over a})^2]^{1\over 2}$
where $n=0,1,2,...$.  Upon combining the
result obtained with the corresponding result for for the interval $a\leq z\leq
L$ where $L\gg a$ is the $z$-coordinate of a third conducting plane, a finite
cutoff independent result $F/A=\pi^2/ 240 a^4$ [4] is obtained for the
Casimir pressure on the plate at $z=a$.

A considerably more elegant approach to this problem is that of Brown and
Maclay [5] who employ an image method to calculate $\langle 0|T^{\mu\nu}(x)
|0\rangle$.  Thus they showed that the photon propagator in the presence
of conducting planes at $z=0$ and $z=a$ could be expressed in terms of an
infinite sum over the usual (i.e., $-\infty <z<\infty$) photon propagator with
the $z$-coordinate of each term in the sum displaced by an even multiple of
$a$.  Since the stress tensor for the electromagnetic case is given by
\begin{equation}
T^{\mu\nu}(x)= F^{\mu\alpha} F^{\nu}_{\;\alpha}-{1\over4}
g^{\mu\nu}F^{\alpha\beta}F_{\alpha\beta}
%\eqno(1)
\end{equation}
where
$$F^{\mu\nu}(x)=\partial^{\mu}A^{\nu}(x)-\partial^{\nu}A^{\mu}(x),$$
it follows that upon taking appropriate derivatives with respect to the
propagator arguments $x$ and $x'$ and invoking the limit $x \rightarrow x'$ a
formal expression can be obtained for the vacuum expectation value of the
stress tensor.  On the basis of covariance arguments together with the
divergence and trace free property of $T^{\mu\nu}(x)$ it was then found in
ref. 5 that
\begin{equation}
\langle 0|T^{\mu\nu}(x)|0\rangle =({1\over 4}g^{\mu\nu}-\hat{z}^{\mu}
\hat{z}^{\nu})({1\over 2\pi^2a^4})\sum_{n=1}^{\infty}n^{-4}
%\eqno(2)
\end{equation}
where $\hat{z}^{\mu}$ is the unit vector (0,0,1,0) in the $z$-direction normal
to the conducting planes.

However, there is some reason to question whether this approach has adequately
dealt with the divergences which invariably occur in Casimir calculations.  One
notes in particular that the result (2) is obtained only after an obviously
singular $n=0$ term has been dropped from the sum which occurs in that
equation.  While one can argue as in [5] that such an $a$-independent term can
be freely omitted since it is merely the usual subtraction of the large
$a$ result, it is well to note that the {\em entire} sum over $n$ is required
for a demonstration that the propagator satisfies correct boundary conditions
at $z=0,a$.  Moreover, as is shown in  this work, an appropriately regularized
form of (2) does not necessarily allow a separation into cutoff dependent terms
and $a$-dependent terms, in contrast with the result found in [5].  Of
still greater import is the fact that more general regularizations than
those usually considered in this calculation lead to an explicit cutoff
dependence of the Casimir stress, a circumstance which would seem to deny its
physical significance.

To establish the above claims one reverts from the image approach to one based
on expansion of the Green's function in terms of orthogonal functions [6]. To
 this end one notes that the free field propagator in the radiation gauge can
be written as
\begin{eqnarray}
G^{ij}({\bf x-x'},z,z',t-t')
= \sum_{n\lambda}\int {d{\bf k} d\omega \over (2\pi)^3}
e^{-i\omega (t-t')} \nonumber \\
\times
{ A_{n\lambda}^i({\bf k},z) A_{n\lambda}^{j*}
({\bf k},z')\over k^2-{\omega}^2+(n\pi/a)^2-i\epsilon}
e^{i{\bf k}\cdot ({\bf x-x'})}
\end{eqnarray}
%eq 3
where $\lambda =1,2$ refers to the polarization, and spatial coordinates
orthogonal to the $z$-direction are denoted by a boldface notation.  The
eigenfunctions $A_{n\lambda}^i({\bf k},z)$ satisfy the equation
$$\left[{\partial^2\over \partial z^2}+(n\pi/ a)^2\right]
A^i_{n\lambda}({\bf k},z)=0$$

and are given explicitly by
\begin{equation}
A_{n1}^i({\bf k},z)={\overline{k}_i\over |{\bf k}|}({2\over a})^{1\over 2}
sin(n\pi z/a)
%\eqno(4)
\end {equation}
and
\begin{equation}
A_{n2}^i({\bf k},z)={1\over |{\bf k}|\omega_k}
\left(\hat{\bf z}^i{\omega}_k^2+
\hat{{\bf z}}\cdot{\bf \nabla}{\nabla}^i\right)
({2\over a})^{1\over 2}cos(n\pi z/a)
%\eqno(5)
\end{equation}
where $\overline{k}_i \equiv \epsilon^{ij}k_j$ with $\epsilon^{ij}$ being the
usual alternating symbol.  It is important to note that
each eigenfunction $A_{n\lambda}^i({\bf k},z)$ satisfies the boundary
conditions $\hat{{\bf z}}\times {\bf E}=\hat{{\bf z}}\cdot{\bf B}=0$ at
$z=0,a$.  This means that it is possible to introduce a regularization such
that contributions from large values of $|{\bf k}|$ and/or $n$ are reduced
without destroying the validity of the boundary conditions.  This stands in
marked contrast with the image method which has no mechanism for the consistent
suppression of the contributions of higher order reflections.

In order to determine the regularization appropriate to this calculation one
should ideally make reference to the underlying symmetry.  Since the latter
consists of the reflection $z\rightarrow a-z$ and the (2+1) dimensional
Poincar\'{e} group, it is natural to seek to classify regularization schemes
according to representations of the latter.  The usual cutoff method for this
problem invokes a parameter which damps out the large ${\omega}_k$
contributions, an approach which makes no reference to the underlying
Lorentz invariance.
A far more appropriate technique is to generalize this to a cutoff based on a
vector ${\sigma}^{\mu}$ in (2+1) dimensions as well as a scalar cutoff
$\Sigma$ which can be used to suppress large values of the (2+1) dimensional
invariant $E^2-{\bf P}^2$ where $E$ and ${\bf P}$ are respectively the energy
and momentum operators associated with this (2+1) dimensional subspace.
Clearly, the credibility  of the Casimir effect requires that the result be
independent of the relative importance of these two competing cutoffs.

The calculation proceeds by noting that since the limit $x\rightarrow x'$ is to
be taken symmetrically at some point, it is appropriate to use only the
imaginary part of the propagator.  An appropriately regularized version of this
function can be inferred from Eq.(3) to be [7]
\begin{eqnarray}
&I&mG_{\sigma,\Sigma}^{ij}(x,x')=\pi\sum_{n\lambda}\int{d^3k\over
(2\pi)^3}\delta(k^2 +(n\pi/a)^2) \nonumber \\
& & \times
A_{n\lambda}^i({\bf k},z)A_{n\lambda}^{j*}({\bf k},z')
e^{ik^{\mu}(x-x')_{\mu}}
%\nonumber \\
% & & \times
e^{\sigma_{\mu}k^{\mu}\epsilon (k^0)}e^{\Sigma(-k^2)^{1\over2}}
\end{eqnarray}
% eq 6
where $\epsilon (k^0)$ is the alternating function and a summation convention
convention has been introduced in the Lorentz invariant subspace.  Note that
since both $\sigma^{\mu}$ and $k^{\mu}$ are three vectors in that space, they
satisfy the orthogonality conditions $\hat{z}^\mu \sigma_\mu =\hat{z}^{\mu}
k_{\mu}=0$ [8].   In addition it is clearly necessary to impose
$\overline\sigma^2 \equiv -\sigma^\mu \sigma_\mu >0$ and $\sigma^0>0$ in
order that this propagator exist.  It will subsequently be found that its
existence also requires  $\Sigma<\overline\sigma$.

To proceed one uses the regularization (6) and the form of the stress tensor
(1).  When used in conjunction with the eigenfunctions (4) and (5) the vacuum
expectation value of the regularized stress tensor can be determined.  With
some effort this is found by straightforward calculation to yield the
coordinate independent result
\begin{eqnarray*}
\langle 0 & | & T^{\mu\nu}|0\rangle  =  {2\pi\over a}\sum_{n=0}^{\infty}
\int { d^3k \over (2\pi)^3}
\delta(k^2 + (n\pi/a)^2)e^{\sigma_\mu k^\mu\epsilon (k^0)} \nonumber \\
& & \times e^{\Sigma n\pi/a}
\left[ k^\mu k^\nu +\hat{z}^\mu \hat{z}^\nu (n\pi/a)^2 \right]
\end{eqnarray*}
which is manifestly both symmetric and traceless. It can be more usefully
written as
\begin{eqnarray}
\langle 0 &| &T^{\mu\nu}|0\rangle = {1\over a}\sum_{n=0}^{\infty}e^{\Sigma
n\pi/a} \nonumber \\
& & \times\left( {\partial\over \partial \sigma_\mu}
{\partial\over \partial\sigma_\nu}
- \hat{z}^\mu \hat{z}^\nu {\partial^2\over \partial\sigma^{\alpha}
\partial\sigma_{\alpha}}\right) \Delta^{n\pi/a}(-i\sigma)
\end{eqnarray}
%eq 7
where $\Delta^{n\pi/a}(x)$ is the (2+1) dimensional function
$$\Delta^{n\pi/a}(x)=2\pi\int{d^3k\over (2\pi)^3}e^{ikx\epsilon (k^0)}
\delta (k^2 +(n\pi/a)^2)$$
for a particle of mass $n\pi/a$.  Since this is an $O(2,1)$ scalar,
$\Delta^{n\pi/a}(-i\sigma)$ is a function of only the invariant
$\overline{\sigma}$ which has the explicit form
$$\Delta^{n\pi/a} (-i\sigma) = {1\over 2\pi\overline{\sigma}}
e^{-\overline{\sigma}n\pi/a}.$$

The insertion of this result into Eq.(7) clearly implies that the sum over $n$
exists only for the case that $\Sigma <\overline{\sigma}$ as previously
stated.  Upon performing the summation over $n$ it follows that
$$
\langle 0|T^{\mu\nu}|0\rangle=
({\partial\over \partial\sigma_\mu \partial\sigma_\nu}-\hat{z}^{\mu}
\hat{z}^{\nu}{\partial^2\over \partial\sigma^{\alpha}\partial\sigma_{\alpha}})
F(\overline{\sigma},\Sigma)
$$
where
$$F(\overline{\sigma},\Sigma) = {1\over 2\pi a\overline {\sigma}}
{1\over 1-e^{(\Sigma - \overline{\sigma}) \pi/a}}.$$
One now performs the usual expansion of the denominator of this expression,
discarding terms which give no contribution in the limit of vanishing cutoff,
thereby obtaining
$$F(\overline{\sigma},\Sigma)\rightarrow
[{1\over 2\pi^2}{1\over
\overline{\sigma}}{1\over \overline{\sigma}-\Sigma}+{1\over 4\pi
a\overline{\sigma}}(1-{\Sigma\pi\over 6a})-{(\overline{\sigma}-\Sigma)^3\pi^2
\over 1440\overline{\sigma}a^4}].$$
Upon performing the derivatives and rearranging terms there finally results [9]
\begin{eqnarray*}
\langle 0 &| &T^{\mu\nu}|0\rangle
 = \left[ g^{\mu\nu}+3{\sigma^{\mu}\sigma^{\nu}\over
\overline{\sigma}^2}-\hat{z}^{\mu}\hat{z}^{\nu} \right] \left\{
{1\over 4\pi a \overline{\sigma}^3}(1-{\Sigma\pi\over 6a}) + \right.
\nonumber \\
& & \left.
{(2\overline{\sigma}-\Sigma)(\overline{\sigma}-\Sigma)
+{2\over 3}\overline{\sigma}^2\over 2\pi^2\overline{\sigma}^3
(\overline \sigma-\Sigma)^3}
+{\pi^2\over 1440a^4}
{\Sigma\over \overline{\sigma}}
\left( {\Sigma^2\over \overline{\sigma^2}}-1\right) \right\}
 \nonumber \\
& & + ({1\over 4}g^{\mu\nu} -\hat{z}^{\mu}\hat{z}^{\nu})
\left[(1-{\Sigma\over \overline{\sigma}}){\pi^2\over 180a^4}
-{4\over 3\pi^2}
{1\over \overline{\sigma}}
{1\over (\overline{\sigma}-\Sigma)^3}\right].
\end{eqnarray*}
If (following [5]) one subtracts the $a\rightarrow \infty$ result, this reduces
to the more tractable form
\begin{eqnarray*}
\langle 0 & | &\overline{T}^{\mu\nu}|0\rangle =
\left[ g^{\mu\nu} +
3 {\sigma^{\mu}\sigma^{\nu}\over \overline{\sigma}^2}
-\hat{z}^{\mu}\hat{z}^{\nu}\right] \nonumber \\
& & \times\left\{ {1\over 4\pi a\overline{\sigma}^3}
\left(1-{\Sigma\pi\over 6a} \right)
+ {\pi^2\over 1440a^4} {\Sigma\over \overline{\sigma}}
\left( {\Sigma^2\over \overline{\sigma}^2}-1 \right) \right\} \nonumber \\
& & + \left( {1\over 4}g^{\mu\nu}-\hat{z}^{\mu}\hat{z}^{\nu}\right)
\left( 1-{\Sigma\over \overline{\sigma}}\right) {\pi^2\over 180a^4}
\end{eqnarray*}
where an overbar notation has been used to denote this subtraction.  It is
noteworthy that even this removal of the large $a$ result does not lead
to regularization independent results, a fact which has been remarked upon
earlier.

Of particular interest to Casimir calculations are the stress components
$\langle 0|\overline{T}^{33}|0\rangle$ and the energy density per unit area
${\cal E}\equiv a\langle 0|\overline{T}^{00}|0\rangle$ which are given by
\begin{equation}
\langle 0|\overline{T}^{33}|0\rangle = -{\pi^2\over 240a^4}
(1-{\Sigma\over \overline{\sigma}})
\end{equation}
and
\begin{eqnarray}
{\cal E} & = & - {\pi^2\over 720a^3} \left\{
 \left( 1-{\Sigma\over \overline{\sigma}}\right) -
{3\sigma_0^2-\overline{\sigma}^2\over 2\overline{\sigma}^2} \right.\nonumber \\
& &\quad \left.\times \left[ {\Sigma\over \overline{\sigma}}
\left( {\Sigma^2\over \overline{\sigma}^2}-1 \right)
- {30\Sigma a^2 \over \pi^2 \overline{\sigma}^3} \right] \right\} +
\nonumber \\
 & & (a \;{\rm independent \; terms})
\end{eqnarray}
% eq 9
respectively.
It is striking that each of these terms retains a significant
dependence on the cutoff details.  In addition the usual relation assumed (as
in [5]) to hold between ${\cal E}$ and the stress components, namely
\begin{equation}
\langle 0|\overline{T}^{33}|0\rangle =-{\partial\over \partial a}{\cal E},
\end{equation}
is manifestly contradicted by Eqs.(8) and (9) in agreement with
results found earlier in the context of the Casimir energy of a sphere [6].  It
is significant that the relation (10) asserts an equality of the vacuum stress
$\langle 0|\overline {T}^{33}|0\rangle$ which transforms under $O(2,1)$ as a
scalar while the right hand side transforms as the $\mu=\nu=0$ component of a
symmetric tensor under this group.  Finally, note should be made of the
fact that Eq.(9) predicts an additional Casimir force proportional to the
divergent indeterminate form $\Sigma/a^2\overline{\sigma}^3$.

To reinforce the conclusions reached here in the case of parallel plates it is
useful to consider also the case of the conducting sphere, the only other
geometry in three dimensions which has proved amenable to exact calculation
[10].  This case was first solved by Boyer [11] and subsequently verified by a
number of authors [12-15].  Following reasoning similar to that of the parallel
plate case note is made of the fact that the unbroken symmetry in this case
consists of time translation and rotational invariance.  Thus the natural
cutoff parameters in this problem should refer to the energy and angular
momentum.  The former is the standard one and is well known to give cutoff
independent results.  It will be the goal here to examine the situation which
occurs when a combination of these two is considered.

This is most economically achieved by reference to [15] which provides a useful
separation of the Casimir energy into a finite part and one which requires
regularization.  Thus one writes for a sphere of radius $a$
$$E_c=E_{fin}+E_{\sigma}$$
where $E_c$, $E_{fin}$, and $E_{\sigma}$ are respectively the total, the
regularization independent part, and the formally divergent parts of the
Casimir energy.  The quantity $E_{\sigma}$ is given by
\begin{eqnarray*}
E_{\sigma}& = & \quad {1\over 4\pi a}\sum_{l=1}^{\infty} Re
e^{-i\phi}\int_0^{\infty}dy
exp(-i\nu\sigma ye^{-i\phi}) y{d\over dy} \nonumber \\
& & \times (1+y^2e^{-2i\phi})^{-3}
\end{eqnarray*}
where $\nu = l+{1\over 2}$, $0<\phi <{\pi\over 2}$, and $\sigma$ is a
dimensionless cutoff used to suppress the high frequency modes.  Upon choosing
a secondary cutoff of the form $e^{-\Sigma \nu}$  it is readily found that
$\Delta E_{\sigma}$ (the {\em change} induced in $E_{\sigma}$ in the limit of
small cutoff) is given by
$$\Delta E_{\sigma} = - {3\Sigma\over 2\pi a\sigma^2}
\int_0^{\infty} dy {y^2 \over {(1+y^2)}^4}
{1\over y^2 + (\Sigma^2/\sigma^2)}.$$
This is evaluated to yield
$$\Delta E_{\sigma}=-{3\over 64 a\sigma}{\Sigma\over
(\Sigma+\sigma)^4}[{\Sigma}^2+4\sigma\Sigma +5{\sigma}^2],$$
a result which displays yet again the cutoff dependence of the Casimir effect
for a more general choice of regularization.  It may be noted that aside from
confirming the vanishing of $\Delta E_{\sigma}$ for $\Sigma =0$, this result
shows that $\Delta E_{\sigma}$ diverges for $\sigma\rightarrow 0$ with all
intermediate values obtained for finite $\Sigma/ \sigma$.

In this work it has been shown that the Casimir effect is, prevailing opinion
notwithstanding, highly dependent on the particular form of regularization
employed for the extraction of the force.  As remarked earlier as well as
in ref.[6] the recent experiments which have seemed to many to provide the long
awaited precision verification of this highly subtle effect are not based upon
rigorous mathematical calculation.  While the parallel plate Casimir experiment
is fraught with difficulties beyond the ken of this author, it would seem that
the successful completion of such experiments would be invaluable for purposes
of setting to rest some of the issues which have been raised in this work.

Finally, it would be remiss not to mention in some way the very extensive work
on the calculation of Casimir forces using the technique of zeta function
regularization [16].  Historically, the successes of the Casimir approach in
dealing with the parallel plate geometry and the sphere were obtained using
conventional field theoretical subtraction procedures.  Specifically, it was
noted that only changes {\em relative} to the vacuum could be considered
observable and it was therefore totally consistent to perform subtractions
relative to the $a\rightarrow \infty$ vacuum.  However, this step did not
succeed in allowing one to obtain finite and observable results in more general
applications.  Eventually it was realized, however, that the application of
zeta function regularization to such problems could yield finite results for
some fairly general cases while at the same time agreeing with those obtained
in the few instances in which more conventional subtractions could be
successfully applied.  This work makes no claim to having established any
inconsistencies in the derivation of finite results for the Casimir effect
when those efforts are based on the twin axioms of vacuum energy {\em and} zeta
function regularization.  Rather, the calculations presented here establish
that the Casimir effect is generally cutoff dependent and hence incapable of
being reliably determined whenever such calculations are performed using
conventional (i.e., physically plausible) subtraction procedures.

\acknowledgments

This work is supported in part by the U.S. Department of Energy Grant
No.DE-FG02-91ER40685.


\begin{references}
\bibitem{ } The units employed here are such that $\hbar=c=1$ with the
signature of the metric being (1,1,1,-1).
\bibitem{ } Equivalence to the usual regularization would require that
$\sigma_i=\Sigma =0.$
\bibitem{ }
It is of interest to note that from the metric $g^{\mu\nu}$ and the vectors
$\sigma^\mu$ and $\hat{z}^\mu$ one can form three second rank tensors
which subsequently reduce to two when the tracelessness condition is applied.
The fact that there is no term of the form $(\sigma^\mu\hat{z}^\nu
+\sigma^\nu\hat{z}^\mu)(a+b \hat{z}^\mu\sigma_\mu)$ is a consequence of the
orthogonality condition $\hat{z}^\mu\sigma_\mu=0$ and the invariance under
$\hat{z}^\mu\rightarrow -\hat{z}^\mu$.
\bibitem{ } This leaves the case of one dimension as the only remaining example
of a Casimir effect calculation which is regularization independent.  In that
application it could hardly be otherwise since the leading singularity is
proportional to the square of an inverse cutoff parameter which is necessarily
removed by subtraction of the $a\rightarrow \infty$ vacuum.  In addition the
next to leading term gives no effect since at most it could contribute a
divergent $a$ independent term to $\cal E$, thereby leaving a finite cutoff
independent remainder to contribute to the force.  In (3+1) dimensions,
however, there are simply too many too many divergences, too many cutoff
parameters, and too few physically reasonable subtractions to obtain a finite
cutoff independent result.
\end{references}
\end{document}